\documentclass[prl,twocolumn,superscriptaddress,showpacs,floatfix]{revtex4}
\usepackage{graphicx}
\usepackage{amssymb}
\usepackage{epstopdf}
\usepackage{amsmath}
\usepackage{amstext}
\usepackage{latexsym}
\usepackage[matrix,frame,arrow]{xypic}

\newcommand{\beq}{\begin{equation}}
\newcommand{\eeq}{\end{equation}}
\newcommand{\barr}{\begin{eqnarray}}
\newcommand{\earr}{\end{eqnarray}}
\newcommand{\vett}[1]{{\bf #1}}
\newcommand{\ket}[1]{\left\vert#1\right\rangle}

\newcommand{\abs}[1]{\left\vert #1 \right\vert}
\newcommand{\Ham}{\mathcal H}
\newcommand{\op}[1]{\hat{#1}}

\begin{document}

\title{Phase diagram of spin-1 bosons on one-dimensional lattices}

\author{Matteo Rizzi}
\affiliation{ NEST- INFM \& Scuola Normale Superiore, Piazza dei
Cavalieri 7 , I-56126 Pisa, Italy}
\homepage{qti.sns.it}

\author{Davide Rossini}
\affiliation{ NEST- INFM \& Scuola Normale Superiore, Piazza dei
Cavalieri 7 , I-56126 Pisa, Italy}
\homepage{qti.sns.it}

\author{Gabriele De~Chiara}
\affiliation{ NEST- INFM \& Scuola Normale Superiore, Piazza dei
Cavalieri 7 , I-56126 Pisa, Italy}
\homepage{qti.sns.it}

\author{Simone Montangero}
\affiliation{ NEST- INFM \& Scuola Normale Superiore, Piazza dei
Cavalieri 7 , I-56126 Pisa, Italy}
\homepage{qti.sns.it}

\author{Rosario Fazio}
\affiliation{ NEST- INFM \& Scuola Normale Superiore, Piazza dei
Cavalieri 7 , I-56126 Pisa, Italy}
\homepage{qti.sns.it}

\date{\today}

\begin{abstract}
Spinor Bose condensates loaded in optical lattices have a rich phase diagram
characterized by different magnetic order.
Here we apply the Density Matrix Renormalization Group to accurately determine
the phase diagram for spin-1 bosons loaded on a one-dimensional lattice.
The Mott lobes present an even or odd asymmetry associated
to the boson filling.
We show that for odd fillings the insulating phase is {\em always}
in a dimerized state. 
The results obtained in this work are also relevant for the determination
of the ground state phase diagram of the $S=1$ Heisenberg model
with biquadratic interaction. 
 
\end{abstract}

\pacs{03.75.Mn, 75.10.Pq, 73.43.Nq}

\maketitle

The experimental realization of optical lattices~\cite{general_experiments} has
paved the way to study strongly correlated many-particle systems with
cold atomic gases (see e.g.~\cite{jaksch05,minguzzi03}). The main advantages 
with respect to condensed matter systems lie on the possibility of a precise 
knowledge of the underlying microscopic models and an accurate and relatively 
easy control of the various couplings. Probably one of the most spectacular
experiments in this respect is the observation~\cite{greiner02} of
a Superfluid - Mott Insulator transition previously predicted
in~\cite{jaksch98} by a mapping onto the Bose-Hubbard model~\cite{fisher89}. 

More recently the use of far-off-resonance optical traps has opened the
possibility to study spinor condensates~\cite{stamperkurn00}. Spin effects are 
enhanced by the presence of strong interactions and small occupation number, 
thus resulting in a rich variety of phases with different magnetic 
ordering. For spin-1 bosons it was predicted that the Mott 
insulating phases have nematic singlet~\cite{demler02}
or dimerized~\cite{yip03} ground state depending on the mean occupation
and on the value of the spin exchange. 
Since the paper by Demler and Zhou~\cite{demler02} several works have 
addressed the properties of the phase diagram of spinor condensates trapped in 
optical lattices~\cite{imambekov03,zhou03,duan03,svidzinsky03,imambekov04,snoek04}.
The increasing attention in spinor optical lattices has also 
revived the attention on open problems in the theory of quantum magnetism.
The spinor Bose-Hubbard model, when the filling corresponds to one boson per 
site, can be mapped onto the $S=1$ Heisenberg model with biquadratic
interactions which exhibits a rich phase diagram including
a long debated nematic to dimer quantum phase
transition~\cite{chubukov91,xian93,fath95,kawashima02,tsuchiya04,porras05,lauchli03,garcia-ripoll04}.

Up to now the location of the phase boundary of the spinor Bose-Hubbard model
has been determined by means of mean-field and strong coupling approaches.
A quantitative calculation of the phase diagram is however still missing.
This might be particularly important in one dimension where non-perturbative
effects are more pronounced. This is the aim of this Letter.
We determine the location of the Mott lobes showing the even/odd asymmetry
in the spinor case discussed in~\cite{demler02}.
We then discuss the magnetic properties of the first lobe, concluding that
it is always in a dimerized phase. 

The effective Bose-Hubbard Hamiltonian, appropriate for $S=1$ bosons, is
given by
\barr 
   \op{\Ham} & = & \frac{U_{0}}{2} \sum_i \op{n}_i (\op{n}_i - 1) + 
   \frac{U_{2}}{2} \sum_i \left(\op{\vett{S}}^2_i - 2 \op{n}_i \right)
   - \mu \sum_i \op{n}_i \nonumber \\
   & - & t \sum_{i,\sigma} \left( 
   \op{a}^{\dag}_{i, \sigma} \op{a}_{i+1, \sigma} + 
   \op{a}^{\dag}_{i+1, \sigma} \op{a}_{i, \sigma}
   \right) \;.
\label{eq:BHM}
\earr
The operator $\op{a}^{\dag}_{i, \sigma}$ creates a boson in the lowest Bloch
band localized on site $i$ and with spin component $\sigma$
along the quantization axis:
$
  \op{n}_i = \sum_{\sigma} \op{a}^{\dag}_{i, \sigma} \op{a}_{i, \sigma}
$ and 
$
  \op{\vett{S}}_i = 
  \sum_{\sigma,\sigma'} \op{a}^{\dag}_{i, \sigma}
  \vett{T}_{\sigma,\sigma'} \op{a}_{i, \sigma'}
$
are the total number of particles and the total spin on site $i$
($\op{\vett{T}}$ are the spin-1 operators). Atoms residing on the same lattice 
site have identical orbital wave function and their spin function must be 
symmetric. This constraint imposes that $S_{i} + n_{i}$ must be even. 
The uniqueness of the completely symmetric state with fixed spin and number 
makes it possible to denote the single site states with
$\ket{n_i, S_i, S^z_i}$. The coupling constants, which obey the constraint
$-1 < U_2/U_0 < 1/2$,
can be expressed in terms of the appropriate Wannier 
functions~\cite{imambekov03}.
$U_0$ is set as the energy scale unit: $U_0=1$.
We discuss only the anti-ferromagnetic case ($0<U_{2}<1/2$).

In the absence of spin dependent coupling a qualitative picture of the phase
diagram can be drawn starting from the case of zero hopping ($t=0$).
The ground state is separated from any excited state by a finite energy gap. 
For finite hopping strength, the energy cost to add or remove
a particle $\Delta E_\pm$ (excitation gap) is reduced and at a critical
value $t_c^\pm (\mu)$ vanishes.
This phase is named the Mott insulator.
For large hopping amplitudes the ground state is a globally coherent
superfluid phase.
When $U_2$ is different from zero, states with lowest 
spins, compatible with the constraint $n_i+S_i=\mbox{even}$, are favoured. 
This introduces an even/odd asymmetry of the lobes: the amplitude of 
lobes with odd filling is reduced as compared with the lobes corresponding
to even fillings~\cite{demler02}.
In the first lobe the extra energy required to have two particles
on a site (instead of one) is $1 +2 U_2- \mu$, thus lowering the
chemical potential value where the second lobe starts.
On the other hand, having no particles on a site gives no gain due to
spin terms, accounting for the nearly unvaried bottom boundary of the lobe.

In order to determine the phase diagram of Eq.(\ref{eq:BHM}) we use
the finite-size Density Matrix Renormalization Group (DMRG) with open boundary 
conditions~\cite{white92-95}.
The strategy of the DMRG is to construct a portion of the system
(called the system block) and then recursively enlarge it, until
the desired system size is reached.
At every step the basis of the corresponding Hamiltonian is truncated, so that
the size of the Hilbert space is kept manageable as the physical system grows.
The truncation of the Hilbert space is performed by retaining the eigenstates
corresponding to the $m$ highest eigenvalues of the block's reduced
density matrix.

The DMRG has been employed, for the spinless case, in~\cite{kuhner98,kuhner00}.
The presence of the spin degree of freedom makes 
the analysis considerably more difficult. In the numerical calculations the 
Hilbert space for the on-site part of the Hamiltonian is fixed by imposing a 
maximum occupation number $\overline{n}_{max}$.
As the first lobe is characterized by an insulating phase with $n=1$ particle
per site we choose $\overline{n}_{max} = 3$ in this case;
the dimension of the Hilbert space per site becomes $d=20$.
We have checked, by increasing the value of  $\overline{n}_{max}$,
that this truncation of the Hilbert space is sufficient to compute
the first lobe. In each DMRG iteration we keep up to $m=300$ states
in order to guarantee accurate results. The numerical calculations of the
second lobe ($n=2$ particles per site) have been performed with
$\overline{n}_{max} = 4$ (which corresponds to $d=35$).

\begin{figure}
  \begin{center}
    \includegraphics[scale=0.35]{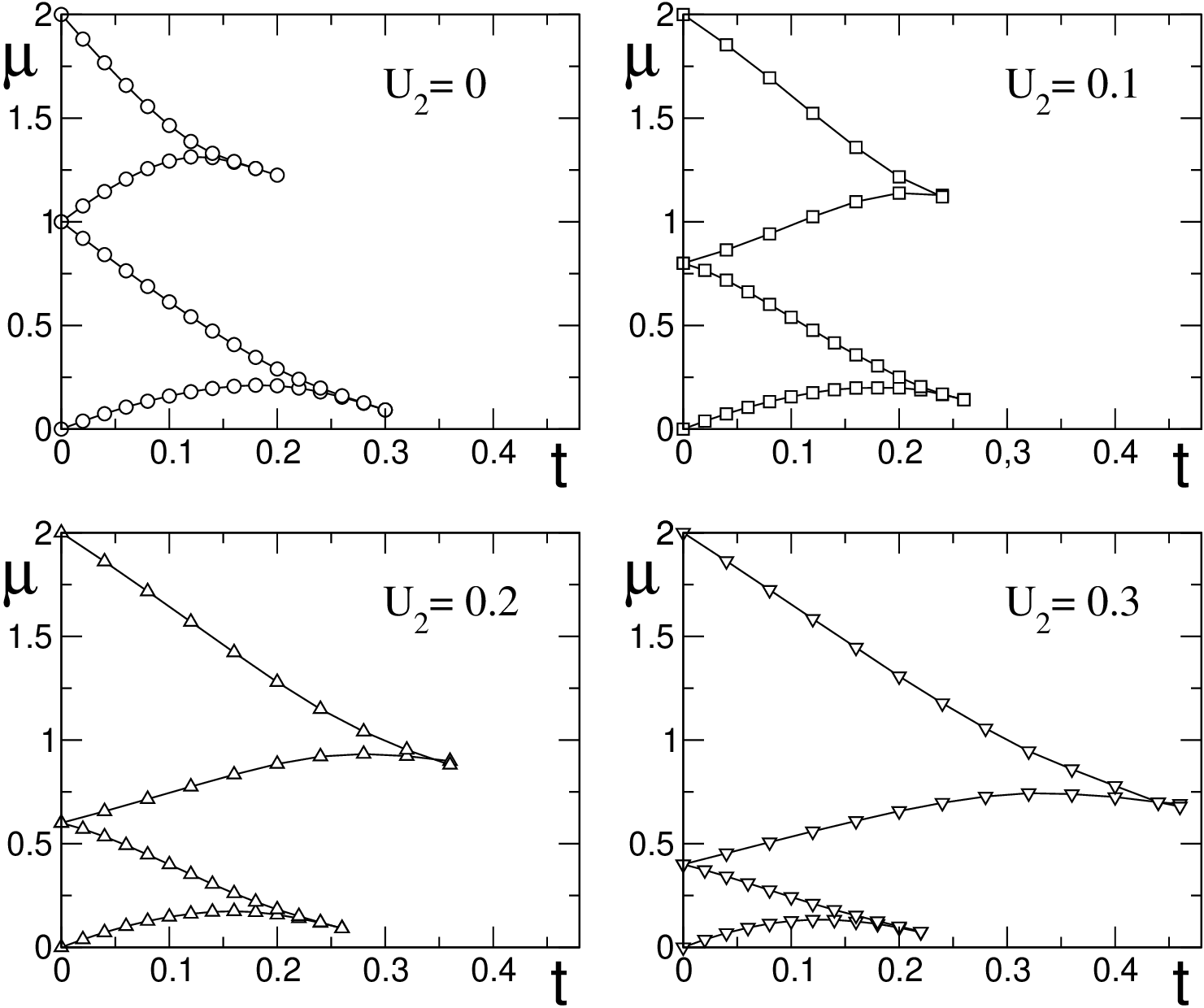}
    \caption{Phase diagram for the first two lobes
      of the 1D Bose-Hubbard spin 1 model with
      nearest-neighbor interactions.
      The different panels correspond to different values of $U_2$.
      The curves for $U_2=0$ coincide with the first two lobes for the
      spinless model computed in Refs.\protect\cite{kuhner98,kuhner00}.}
    \label{fig:lobes}
  \end{center}
\end{figure}

{\em Phase Diagram } - 
In the insulating phase the first excited state is separated
by the ground state by a Mott gap. In the limit of zero hopping
the gap is determined by the extra energy $\Delta E_\pm$ needed
to place/remove a boson at a given site. The finite hopping renormalizes the
gap which will vanish at a critical value. Then the system becomes superfluid.
This method has been employed for the spinless case by Freericks and Monien
\cite{freericks96}, and in~\cite{kuhner98,kuhner00} where it was combined
with the DMRG. Here we use it for the spinor case.
Three iterations of the DMRG procedure are performed, with projections 
on different number sectors; the corresponding ground states give the desired
energies $E_0$, $E_\pm = E_0 + \Delta E_\pm$.
As target energies we used those obtained by the mapping of the Bose-Hubbard
system into effective models as described in~\cite{imambekov03}.
We considered chains up to $L=128$ sites for the first lobe, and $L=48$ for
the second lobe.
The extrapolation procedure to extract the asymptotic values was obtained
by means of linear fit in $1/L$, as discussed in~\cite{kuhner00}.
A comparison with a quadratic fit shows that $O(1/L^2)$ corrections are
negligible on the scale of Fig.\ref{fig:lobes}.

The plot of the phase diagram in the $(\mu,t)$ plane for different
values of the spin coupling $U_2$ is shown in Fig.\ref{fig:lobes}.
The first lobe tends to reduce its size on increasing the spin coupling;
in particular the upper critical chemical potential at $t=0$ is
$\mu_c^+ (0)=1-2 U_2$, while the $t^*$ value of the hopping strength
over which the system is always superfluid is suppressed as $U_2$
increases.
On the other hand, the second lobe grows up when $U_2$ increases.
This even/odd effect, predicted in~\cite{demler02},
is quantified in Fig.\ref{fig:lobes}.

{\em Magnetic properties of the first Mott lobe} - The first lobe 
of the spinor Bose lattice has a very interesting magnetic structure.
In presence of small hopping $t$ boson tunneling processes induce 
effective pairwise magnetic interactions between the spins, described by
Hamiltonian~\cite{imambekov03}:
\begin{equation}
\label{eq:mapping}
\op{\Ham}_{\mathrm{eff}} = 
\kappa \sum_{\langle i j \rangle} \left[ \cos \theta \,
( \op{\vett{S}}_i \cdot \op{\vett{S}}_j )
+ \sin \theta \, ( \op{\vett{S}}_i \cdot \op{\vett{S}}_j )^2
\right]
\end{equation}
with
\beq
\tan \theta = \frac{1}{1-2 U_2} \hspace{8mm} \kappa = \frac{2 t^2}{1+U_2}
\sqrt{1+\tan^2 \theta} \;\; .
\eeq
The absence of higher order terms, such as
$( \op{\vett{S}}_i \cdot \op{\vett{S}}_j )^{3}$, is due to the fact
that the product of any three spin operators can be expressed via lower order
terms. In the case of anti-ferromagnetic interaction in Eq.(\ref{eq:BHM}), 
the parameter $\theta$ varies in the interval $\theta \in [-3/4 \pi, -\pi /2[$.
Because of the form of the magnetic Hamiltonian, each bond tends to 
form a singlet-spin configuration, but singlet states on neighboring bonds 
are not allowed. 
There are two possible ground states that may appear in this situation. 
A nematic state can be constructed by mixing states with total spin 
$S=0$ and $S=2$ on each bond. This construction can be repeated on 
neighboring bonds, thereby preserving translational invariance.
This state breaks the spin-space rotational group $O(3)$, though time-reversal
symmetry is preserved. The expectation value of any spin operator vanishes
($\langle \op{S}^\alpha_i \rangle = 0 , \; \alpha= x,y,z$),
while some of the quadrupole operators have finite expectation values.
The tensor $\mathcal{Q}^{ab} = \langle \op{S}^a \op{S}^b \rangle
- \frac{2}{3} \delta^{ab}$ is a traceless diagonal matrix,
due to invariance under spin reflections.
Since it has two identical eigenvalues ($\langle (\op{S}^x_i)^2 \rangle =
\langle (\op{S}^y_i)^2 \rangle \neq \langle (\op{S}^z_i)^2 \rangle$),
it can be written as
$\mathcal{Q}^{ab} = Q \left( d^a d^b - \frac{1}{3} \delta^{ab} \right)$
using an order parameter
$
\langle \op{Q} \rangle \equiv \langle (\op{S}^z_i)^2 \rangle - \langle
(\op{S}^x_i)^2 \rangle =$
$\frac{3}{2} \langle (\op{S}^z_i)^2 \rangle - 1
$
and a unit vector $\vett{d}= \pm \vett{z}$.
However, since $[ \op{Q},\op{\Ham}_{\mathrm{eff}} ] =0$, it is not possible
to get $Q \neq 0$ in finite-size systems, analogously to what happens
for the magnetization without external field.
Therefore we characterized the range of nematic correlations in
the ground state by coupling this operator to a fictitious ``nematic field'':
$\op{\Ham}_\lambda = \op{\Ham}_{\mathrm{eff}} + \lambda \op{Q}$,
and by evaluating the nematic susceptibility $\chi_{\mathrm{nem}}$
as a function of $L$:
\beq
\label{eq:suscept}
\chi_{\mathrm{nem}} \equiv - \left.
\frac{\mathrm{d}^2 E_0 (\lambda)}{\mathrm{d} \lambda^2}
\right|_{\lambda=0}
= \sum_\gamma \frac{\left| Q_{0,\gamma} \right|^{2}}{E_\gamma - E_0} \, ,
\eeq
where $E_0 (\lambda)$ is the ground energy of $\op{\Ham}_\lambda$, \,
$Q_{0,\gamma}$ is the matrix element between the ground and an excited state
of $\op{\Ham}_{\mathrm{eff}}$ (respectively with energy $E_0$ and $E_\gamma$).

\begin{figure}
  \begin{center}
    \includegraphics[scale=0.35]{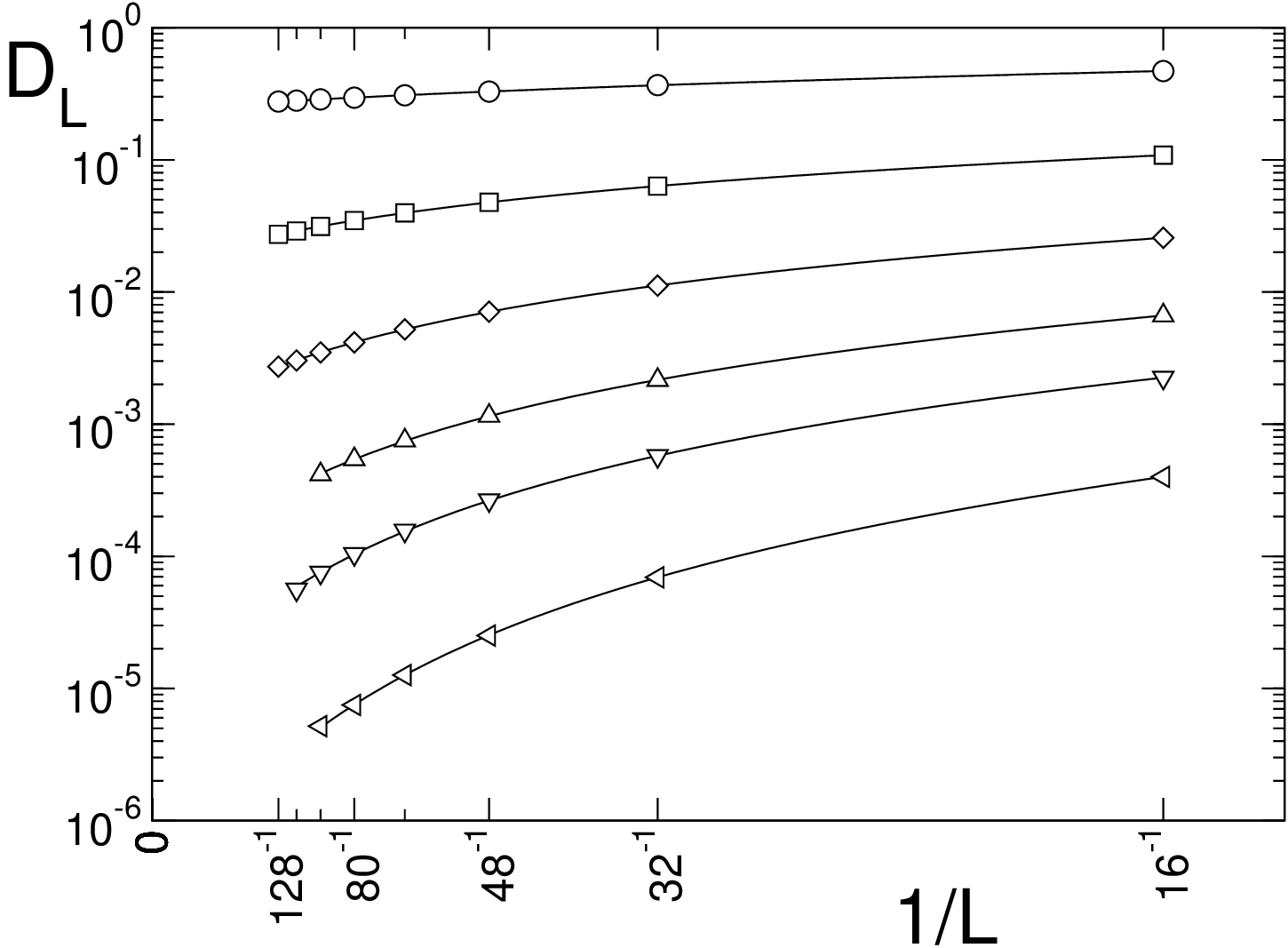}
    \caption{
      Finite-size scaling of $D_L$ for selected
      values of $\theta$: circles ($\theta=-0.65 \pi$),
      squares ($-0.7 \pi$), diamonds ($-0.72 \pi$),
      triangles up ($-0.73 \pi$), triangles down ($-0.735 \pi$),
      triangles left ($-0.74 \pi$).
      In order to extrapolate the order parameter $D$, numerical data
      have been fitted with $D_L = D + c L^{-\alpha}$ (straight lines).
      DMRG simulations are performed with $m \simeq 140$ for
      $\theta > -0.73 \pi$, and $m \simeq 250$ for $\theta \leq -0.73 \pi$.}
    \label{fig:dimer1}
  \end{center}
\end{figure}

On the other hand a possibility to have $SO(3)$ symmetric solution stems
from breaking translational invariance.
Indeed a dimerized solution with singlets on every
second bond satisfy these requirements. Dimerization could be described looking
at the differences in expectation values of pair Hamiltonian
$\op{\Ham}^{(ij)}_{\mathrm{eff}}$ 
on adjacent links ($\op{\Ham}_{\mathrm{eff}}=\sum_{\langle i j \rangle} 
\op{\Ham}^{(ij)}_{\mathrm{eff}}$)~\cite{footnote}. 
The order parameter $D$ reads
\beq
D \equiv \abs{\langle \op{\Ham}^{(i-1,i)}_{\mathrm{eff}} -
\op{\Ham}^{(i,i+1)}_{\mathrm{eff}} \rangle}.
\label{eq:dimer}
\eeq
\begin{figure}
  \begin{center}
    \includegraphics[scale=0.35]{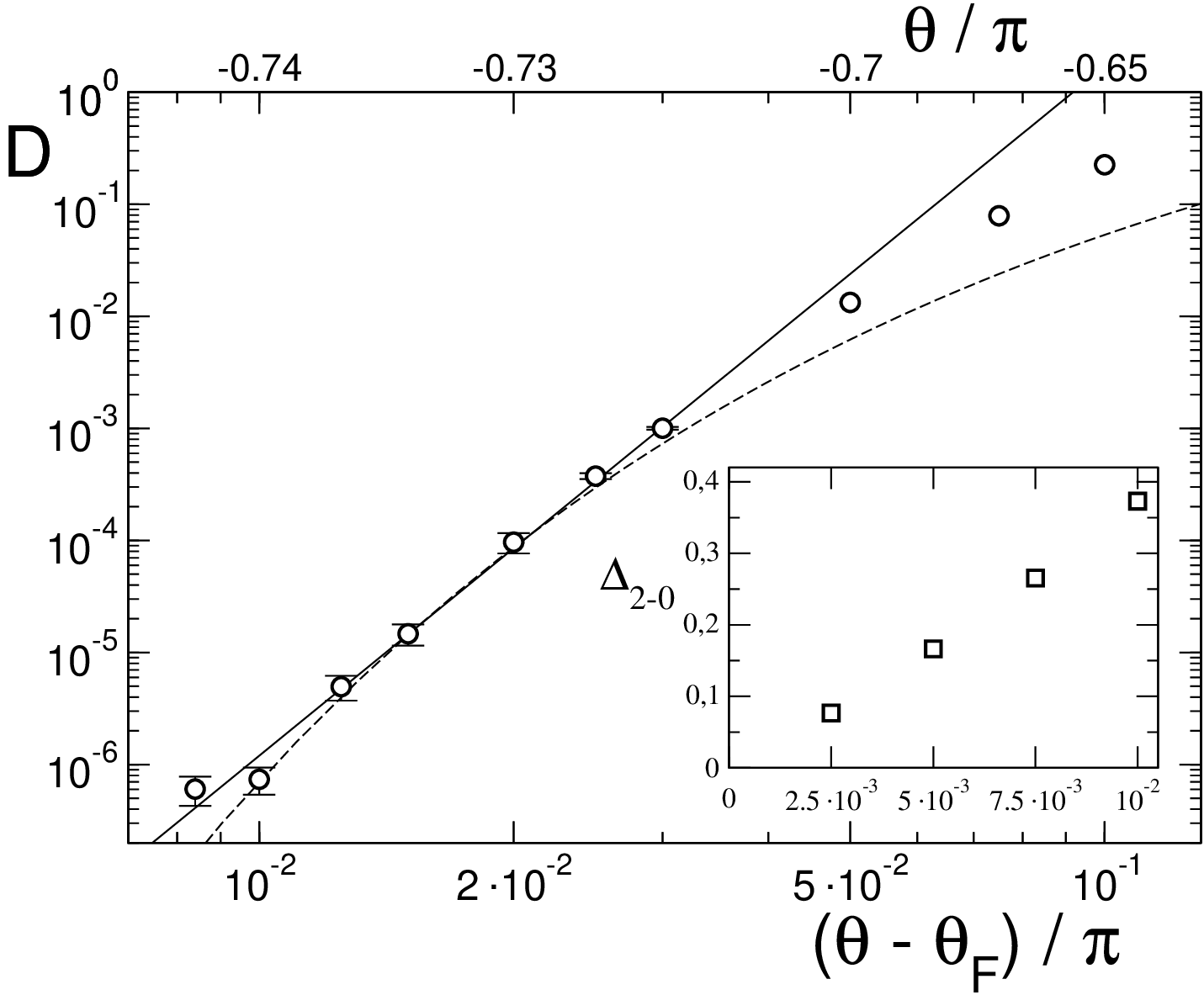}
    \caption{Dimerization order parameter $D$
      near the ferromagnetic boundary:
      solid line shows a power law fit $D \sim (\theta - \theta_F)^{\gamma}$
      of numerical data with $\gamma \simeq 6.15$;
      dashed line shows an exponential law fit
      $D \sim \exp [-a/(\theta - \theta_F)^{-1/2}]$ with $a \simeq 2.91$.
      The linear fit is done over data for $\theta < -0.7 \pi$,
      while the exponential fit is for $\theta \leq -0.73 \pi$.
      DMRG calculations are performed with up to $m \simeq 300$ states.
      Inset: extrapolated scaled gap
      $\Delta_{2-0} =(L-1) (E_2 - E_0)$
      at the thermodynamic limit, close to $\theta_F$.}
    \label{fig:dimer2}
  \end{center}
\end{figure}
It has been proposed~\cite{chubukov91} that a narrow nematic region exists
between the ferromagnetic phase boundary ($\theta_F = -3 \pi /4$, i.e.
$U_2 = 0$) and a critical angle $\theta_C \approx -0.7 \pi$
(i.e. $U_2 \sim 10^{-2}$),
whereas a dimerized solution is favoured in the remaining anti-ferromagnetic
region $\theta_C \leq \theta \leq -\pi/2$. This implies that the dimerization
order parameter D should scale to zero in the whole nematic region.
This possibility has been analyzed in Ref.~\cite{fath95} where it was suggested
that $D$ might go to $0$ exponentially near the ferromagnetic
boundary, making it difficult to detect the effective existence of the nematic
phase. This interesting challenge has motivated numerical investigations with
different methods~\cite{fath95,kawashima02,porras05,lauchli03}.
We present new DMRG results which clarify the magnetic properties
of the first Mott lobe (for sufficiently small hopping) and, consequently,
of the biquadratic Heisenberg chain.

According to our numerical calculation there is {\em no} intermediate
nematic phase, indeed we found a power law decay of the dimerization
order parameter near $\theta_F = -3 \pi /4$.
The simulations of the bilinear-biquadratic model (\ref{eq:mapping}) are less
time and memory consuming than Bose-Hubbard ones, since the local Hilbert space
has a finite dimension $d=3$.
The number of block states kept during the renormalization procedure was chosen
step by step in order to avoid artificial symmetry breaking.
This careful treatment insures that there are no spurious sources of asymmetry
like partially taking into account a probability multiplet.
Here we considered up to $m \simeq 300$ states in order to obtain stable
results.
Raw numerical data are shown in Fig.\ref{fig:dimer1}, where
the finite-size dimerization parameter $D(L)$ is plotted as a function of the
chain length $L$ (see Eq.~\ref{eq:dimer}, and \cite{footnote}).
Finite-size scaling was used to extrapolate to the thermodynamic limit.
After the extrapolation to the $L \to \infty$ limit, see Fig.~\ref{fig:dimer2},
we fitted the dimer order parameter with a power law
\beq
D = \left( \frac{\theta - \theta_F}{\theta_0} \right)^{\gamma}
\label{dimerfit}
\eeq
where $\gamma \sim 6.1502$ and $\theta_0 \sim 0.09177 \, \pi$
(Fig.~\ref{fig:dimer2}, solid line).
We also tried to fit our data by an exponential law
\beq
D = D_0 \, e^{-a/\sqrt{\theta - \theta_F}}
\eeq
as suggested in~\cite{fath95},
with $a \sim 2.911$, $D_0 \sim 9.617$; this fit seems to work for
narrower regions (Fig.~\ref{fig:dimer2}, dashed line), however
from our numerics we cannot exclude an exponential behavior
of $D$ in the critical region.
The dimerized phase thus seems to survive up to the ferromagnetic
phase boundary, independently from the chosen fitting form.
This is also confirmed by the fact that the scaled gap between
the ground state $E_0$ and the lowest excited state $E_2$
(which is found to have total spin $S_{T}=2$) seems not to vanish in the
interesting region $\theta > -0.75 \pi$ (see inset of Fig.~\ref{fig:dimer2}).

\begin{figure}
  \begin{center}
    \includegraphics[scale=0.33]{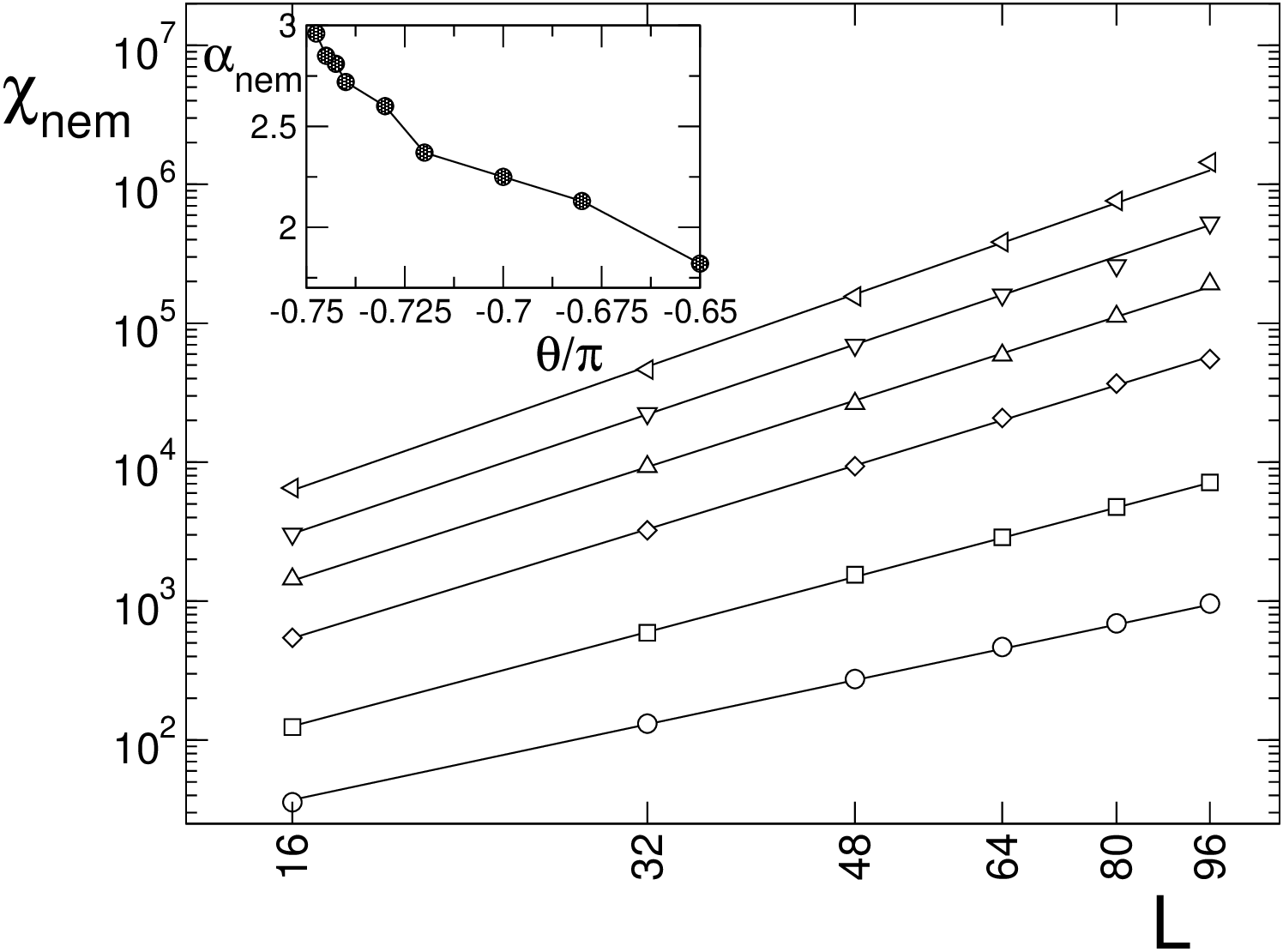}
    \caption{Nematic susceptibility $\chi_{\mathrm{nem}}$ as a function
      of the system size $L$.
      The various symbols refer to different values of $\theta$:
      circles ($\theta=-0.65 \pi$),
      squares ($-0.7 \pi$), diamonds ($-0.73 \pi$),
      triangles up ($-0.74 \pi$), triangles down ($-0.745 \pi$),
      triangles left ($-0.7475 \pi$).
      Straight lines show a power law fit
      $\chi_{\mathrm{nem}} = c L^\alpha$ of numerical data.
      Inset: exponent $\alpha$ as a function of $\theta$.}
    \label{fig:nematic}
  \end{center}
\end{figure}

Moreover we analyzed the 
susceptibility of the chain to nematic ordering $\chi_{\mathrm{nem}}$.
The numerical data, presented in Fig.~\ref{fig:nematic},
show a power law behavior $\chi_{\mathrm{nem}} (L) \propto L^\alpha$
as a function of the system size.
The exponent $\alpha$ (shown in the inset) approaches the value 
$\alpha=3$ as $\theta \to \theta_F$. This can also be confirmed by means of
a perturbative calculation around the exact solution available at $\theta_F$;
indeed one obtains $\left| Q_{0,\gamma} \right|^{2} \sim L^{2}$ and
$(E_\gamma - E_0) \sim L^{-1}$ to be inserted in Eq.~(\ref{eq:suscept}).
The increase of the exponent for $\theta \to \theta _F$ indicates, as suggested
in~\cite{porras05}, that a tendency towards the nematic ordering is enhanced as
the dimer order parameter goes to zero.

{\em Conclusions } - In this Letter we analyzed, by means of a DMRG analysis,
the phase diagram of the one-dimensional spinor boson condensate on an
optical lattice.
We determined quantitatively the shape of the first two Mott lobes,
and the even/odd properties of the lobes. We furthermore discussed the 
magnetic properties of the first lobe. Our results indicate that the
Mott insulator is {\em always} in a dimerized phase. 

This work was  supported by IBM (2005 Faculty award), and by the European 
Community through grants RTNNANO, SQUBIT2.


\begin{thebibliography}{99}
\bibitem{general_experiments}
	B.P. Anderson and M.A. Kasevich, Science {\bf 282}, 1686 (1998);
	F.S. Cataliotti {\it et al}., ibid {\bf 293}, 843 (2001);
	S. Burger {\it et al}., Phys. Rev. Lett. {\bf 86}, 4447 (2001);
	O. Morsch {\it et al}., ibid {\bf 87}, 140402 (2001).
\bibitem{jaksch05}
        D. Jaksch and P. Zoller; Ann. Phys. {\bf 315}, 52 (2005).
\bibitem{minguzzi03}
	A. Minguzzi {\it et al}.,
	Phys. Rep. {\bf 395}, 223 (2004).
\bibitem{greiner02}
	M. Greiner {\it et al}.,
	Nature {\bf 415}, 39 (2002).
\bibitem{jaksch98}
	D. Jaksch {\it et al}.,
	Phys. Rev. Lett. {\bf 81}, 3108 (1998).
\bibitem{fisher89}
	M.P.A. Fisher {\it et al}.,
	Phys. Rev. B {\bf 40}, 546 (1989).
\bibitem{stamperkurn00} 
	For a review of spinor condensates, see D.M. Stamper-Kurn and 
	W. Ketterle, in {\it Coherent Atomic Matter Waves},
	Proceedings of the Les Houches 1999 Summer School, Session LXXII,
	edited by R. Kaiser, C. Westbrook, and F. David
	(Springer, New York, 2001); cond-mat/0005001.
\bibitem{demler02}
	E. Demler and F. Zhou, Phys. Rev. Lett. {\bf 88}, 163001 (2002).
\bibitem{yip03}
	S.K. Yip, Phys. Rev. Lett. {\bf 90}, 250402 (2003).
\bibitem{imambekov03}
	A. Imambekov, M. Lukin, and E. Demler,
	Phys. Rev. A {\bf 68}, 063602 (2003) 
\bibitem{zhou03}
	F. Zhou and M. Snoek, Annals of Physics {\bf 308}, 692 (2003).
\bibitem{duan03}
	L.M. Duan, E. Demler, and M.D. Lukin,
	Phys. Rev. Lett. {\bf 91}, 090402 (2003).
\bibitem{svidzinsky03}
	A.A. Svidzinsky and S.T. Chui,
	Phys. Rev. A {\bf 68}, 043612 (2003).
\bibitem{imambekov04}
	A. Imambekov, M. Lukin, and E. Demler,
	Phys. Rev. Lett. {\bf 93}, 120405 (2004)
\bibitem{snoek04}
	M. Snoek and F. Zhou, Phys. Rev. B {\bf 69}, 094410 (2004). 
\bibitem{chubukov91}
        A.V. Chubukov, Phys. Rev. B {\bf 43}, 3337 (1991).
\bibitem{xian93}
        Y. Xian, J. Phys. Condens. Matter {\bf 5}, 7489 (1993).
\bibitem{fath95}
        G. F\`ath and J. S\`olyom,
        Phys. Rev. B {\bf 51}, 3620 (1995).
\bibitem{kawashima02}
        N. Kawashima, Prog. Theor. Phys. Suppl. {\bf 145}, 138 (2002).
\bibitem{tsuchiya04}
        S. Tsuchiya, S. Kurihara, and T. Kimura,
        Phys. Rev. A {\bf 70} 043628 (2004).
\bibitem{porras05}
        D. Porras, F. Verstraete, and J.I. Cirac,
        cond-mat/0504717.
\bibitem{lauchli03}
        A. L\"auchli, G. Schmid, and S. Trebst,
	cond-mat/0311082.
\bibitem{garcia-ripoll04}
        J.J. Garc\'ia-Ripoll, M.A. Martin-Delgado, and J.I. Cirac,
	Phys. Rev. Lett. {\bf 93}, 250405 (2004).
\bibitem{white92-95}
        S.R. White, Phys. Rev. Lett. {\bf 69}, 2863 (1992);
        Phys. Rev. B {\bf 48}, 10345 (1993).
\bibitem{kuhner98}
        T.D. K\"uhner and H.Monien,
        Phys. Rev. B {\bf 58}, R14741 (1998).
\bibitem{kuhner00}
        T.D. K\"uhner, S.R. White, and H. Monien,
        Phys. Rev. B {\bf 61}, 12474 (2000).
\bibitem{freericks96}
        J. K. Freericks and H. Monien, Phys. Rev. B {\bf 53}, 2691 (1996).
\bibitem{footnote}
	On any finite chain some inhomogeneity exists, thus leading to a
	finite $D_L$ even if $D=0$. Quantitatively, an order parameter $D_L$
	could be defined by evaluating Eq.~(\ref{eq:dimer}) in the middle
	of the finite-size chain.
	The order parameter $D$ has to be extrapolated in the limit
	$D \equiv \lim_{L\rightarrow\infty} D_L$.
\end{thebibliography}
\end{document}